# Strong coupling of Fe-Co alloy with ultralow damping to superconducting co-planar waveguide resonators

I.W. Haygood[1,2], M.R. Pufall[1], E.R.J. Edwards[3], Justin M. Shaw[1], W.H. Rippard[1]

[1]National Institute of Standards and Technology, Boulder, CO, 80305
[2]Department of Physics, University of Colorado, Boulder, CO, 80309
[3]IBM corporation, Albany, NY, 12203

Abstract

We report on the strong coupling between a metallic ferromagnetic $Fe_{75}Co_{25}$ thin film patterned element and a range of superconducting Nb half-wavelength co-planar waveguide (CPW) resonators. By varying the volume of the ferromagnet we demonstrate that the coupling rate scales linearly with the square root of the number of spins and achieve a coupling rate over 700 MHz, approaching the ultrastrong coupling regime. Experiments varying the center conductor width while maintaining constant magnetic volume verify that decreasing the center conductor width increases coupling and cooperativity. Our results show that the frequency dependence of the coupling rate is linear with the fundamental and higher order odd harmonics of the CPW, but with differing efficiencies. The results show promise for scaling planar superconducting resonator/magnetic hybrid systems to smaller dimensions.

Quantum technologies based on hybrid systems where light is strongly coupled to a degree of freedom in a solid state system have the potential to overcome some of the practical engineering challenges limiting large-scale quantum computing.[1] A hybrid system that has been recently investigated is based on the interaction between resonant microwave photons and the collective spin excitations in a magnetic structure.[2] Experiments exploiting the coherent coupling between the two systems have demonstrated strong and ultra-strong coupling regimes, magnetically induced transparency, the Purcell effect, cavity mediated spin-spin coupling, and even potential quantum memories.[3–15] Dissipative coupling has been used to explore frequency attraction and investigate the non-Hermitian physics possible in such systems.[16–22] At millikelvin temperatures, strong coupling between magnons and a qubit, mediated by a cavity, have been used to measure individual magnon and photon numbers.[23,24] Other experiments have demonstrated the potential for optical-to-microwave transduction, a critical technology for quantum information systems (QIS), using magneto-optical coupling between optical and spin-wave modes.[25–28]

Typically, the microwave resonator used in these experiments is a three-dimensional cavity and the magnetic element is a highly polished yttrium iron garnet (YIG, $Y_3Fe_5O_{12}$) sphere, however other geometries have also been explored. YIG is an excellent material due to its low damping, which allows for long magnon lifetimes and microwave cavities with high quality factors (Q) are readily available allowing for long photon lifetimes. While this specific configuration has been used with much success, many promising quantum computing platforms are microfabricated, and it is likely that any scalable quantum system will largely be lithographically defined to efficiently scale, and utilize planar signal lines and resonators such as the co-planar-waveguide (CPW) resonator. Consequently, it is likely beneficial to have a similar lithographically defined system for hybrid magnetic systems. The smallest commercially available YIG spheres are on the order of 200 μm in diameter and would likely be incompatible with such a system.

Planar YIG samples are also available but with the limitation that they must be grown on gadolinium gallium garnet (GGG) to have desirable damping rates.[29] While it may be possible to pattern YIG and fabricate planar resonators, GGG has high microwave losses at low temperatures making it a nonideal substrate.[30] Due to these limitations, two recent papers have investigated coupling between micro-fabricated thin films of Permalloy ($Ni_{80}Fe_{20}$) and a superconducting CPW resonator.[31,32]

In this paper we explore the coupling between superconducting Nb CPW resonators and the ultra-low damping alloy $Fe_{75}Co_{25}$.[33] This alloy shows promise for planar hybrid systems for two main reasons. One, it has a high saturation magnetization $\mu_0 M_s$ and thus a high spin density. Two, the intrinsic damping, $\alpha$, is much lower, about 5 x $10^{-4}$, than many other metallic ferromagnets. Because of these properties, we are able to demonstrate a coupling strength of 707 MHz ± 11 MHz between a half wavelength *(λ/2)* Nb CPW resonator and a $Fe_{75}Co_{25}$ magnetic volume of 192 $\mu m^3$, far into the strong coupling regime for this system. Along with verifying that coupling in this system scales with the square root of the number of spins $\sqrt{N}$, we investigate the change in magnon-photon coupling with increasing resonator frequency and increasing resonator harmonic number. The coupling increases linearly with increasing fundamental frequency of the resonator, as well as with odd harmonic number for a given resonator. However, higher order harmonics couple less efficiently compared to a resonator of equivalent fundamental frequency (shorter physical length). Finally, we show that the coupling increases with the inverse of the resonator center conductor width $1/W_{cc}$ for a given magnetic volume, demonstrating a promising pathway for scaling to smaller magnetic elements.

The resonators were fabricated by first depositing a metallic stack of 5Ti/5Cu/40$Fe_{75}Co_{25}$/5Cu/5Ti, with the preceding numbers indicating the layer thickness in nanometers, on an oxidized Si wafer. A 40 nm ferromagnetic layer was chosen to maximize the magnetic volume to ensure a strong signal for a given patterned area while still maintaining low damping values[33,34]. The magnetic elements were defined using standard optical lithography and ion milling. The patterned magnet is narrower than $W_{cc}$ by 2 μm and is centered on the resonator both laterally and longitudinally, corresponding to an anti-node in the current and maximum RF magnetic field. After patterning the magnetic film, a 185 nm layer of Nb was blanket sputter deposited. The CPW resonators were then patterned using reactive ion etching. Finally contact pads as well as various alignment marks were deposited by a sputtering Au-Pd lift-off process. A representative device is shown in Figure 1(a) along with the ground-signal-ground (GSG) probes used to couple to the resonator.

To investigate the coupling between the defined ferromagnets and the Nb resonators the Fe-Co sample was placed in a cryogenic probe station with base temperature of 3.2 K.[35] Electrical contact was made using a set of GSG probes whose position could be adjusted *in situ* (with approximately 10 μm resolution). This flexibility allowed for the interrogation of the hybrid system by capacitively coupling to the CPW, resulting in a resonant measurement, or by placing the probes in direct contact with the center conductor, resulting in a broadband measurement. For all measurements, a field was applied along the length of the resonator using a set of superconducting coils and the transmission coefficient $S_{21}$, was captured using a vector network analyzer (VNA) calibrated to the input and output ports of the VNA. This in-plane field results in a parabolic change in resonator frequency versus field due to a reduction in the fraction of superconducting electron pairs[36,37]. This is shown in Figure 1(b) for a bare resonator (i.e., one with no magnetic element) demonstrating a resonance frequency, $\omega_r/(2\pi)$ shift of less than 15 MHz at an applied field of 600 mT. Additionally the applied magnetic field results in a small degradation of the resonator quality factor, $Q$, shown in Figure 1(c), with $Q > 2000$ corresponding to a maximum linewidth of $\kappa_r/(2\pi)$ = 3.2 MHz at 8 GHz.

In order to get the resonator mode $\omega_r$ and the ferromagnetic resonance (FMR) mode $\omega_{FMR}$ to interact it is necessary to apply a magnetic field to bring the frequency of the magnetic resonance close to the CPW resonance frequency. It can be shown that for rectangular ferromagnetic prisms the resonance frequency of the mode corresponding to uniform magnetic precession is given by[38,39]

$$\frac{\omega_{FMR}}{2\pi} = \frac{\gamma}{2\pi}\sqrt{(B_{ext} + (N_y - N_z)\mu_0 M_s)(B_{ext} + (N_x - N_z)\mu_0 M_s)}, \qquad (1)$$

where $\gamma$ is the electron gyromagnetic ratio, $B_{ext} = \mu_0 H_{ext}$ the in-plane magnetic flux density, $\mu_0$ the vacuum permeability, and $M_s$ is the saturation magnetization. $N_{x,y,z}$ are the demagnetizing factors for a given axis, with $z$-axis along the resonator length, $x$-axis normal to the length in the plane, and $y$-axis normal to the resonator plane. For high aspect ratio ferromagnets, the demagnetizing factors have a significant effect on $\omega_{FMR}$, compared to blanket thin films, especially at small applied fields.

When the externally applied field brings the two resonances near each other the systems can effectively and coherently transfer energy, and a characteristic avoided crossing occurs. A simple model for the dispersion of the hybrid system is given by two interacting harmonic oscillators and the corresponding eigenfrequencies are[3,40]

$$\omega_\pm = \omega_r + \frac{\Delta}{2} \pm \frac{1}{2}\sqrt{\Delta^2 + 4g^2}, \qquad (2)$$

with $\Delta \equiv \omega_{FMR} - \omega_r$, and $g$ is the coupling rate between the two modes. A typical plot of the transmitted power vs. applied field and frequency is shown in *Figure 2*(a), where the magnetic field was swept from negative to positive. The room-temperature VNA source had a nominal output power of -5 dBm that was attenuated by approximately 20 dB at the sample due to the cabling used in the probe station. The quality factor of the resonator is reduced at low fields likely due to incomplete saturation of the magnetization in the sample leading to the coupling of inhomogeneous magnetic modes to the resonator. The quality factor is restored to a value close to that of a bare resonator at higher fields when the FMR frequency has moved far from the CPW resonance. In order to fit the spectral data to the eigenvalue solution, the peak in the transmission at each applied field is fit to a Lorentzian. This is shown in *Figure 2*(B) and (C). The Lorentzian fit accurately captures the frequency and linewidth; however the data typically have longer tails (whose origins are unknown) compared to the ideal response. With the maximum transmission of the system now established and using Eq. 1 with $\mu_0 M_s$ = 2.45 T, we can now use Eq. 2 to extract the coupling rate for the hybrid system.

The results of the above analysis for two sets of resonators, one with $W_{cc}$ = 50 μm and one with $W_{cc}$ = 10 μm, and with magnetic elements ranging in length from 100 μm to 1000 μm, are shown in Figure 3. All resonators had a nominal resonance frequency of 8 GHz, which was slightly affected by the presence of the magnetic element. Figure 3 shows that for both center conductor widths, the coupling scales linearly with the square root of the magnetic volume. Field-swept FMR measurements performed on a 48 μm wide and 250 μm long sample at 8 GHz had a full-width at half-maximum of $\mu_0 \Delta H$ = 4.6 mT, which resulting in a magnon damping rate[41] $\kappa_m/(2\pi)$ = 156 MHz. In-plane FMR measurements vs. frequency yield a damping parameter, $\alpha$ = 0.005 and inhomogeneous broadening, $\mu_0 \Delta H_0$ = 1.6 mT. While the measured damping is larger than the intrinsic damping,[34] it is still generally consistent, considering that in this experimental setup the measured linewidths can be broadened due to two-magnon scattering along with additional radiative damping due to the coupling to the resonator. It should be noted that independent measurements of $Fe_{75}Co_{25}$ films show $\mu_0\Delta H$ = 1.7 mT, corresponding to $\kappa_m/(2\pi)$ = 58 MHz indicating the importance of measuring patterned magnetic samples at 4K to get accurate total magnon damping rates. Nonetheless, using the worst-case values for magnon and resonator damping we reach a cooperativity, $C = g^2/(\kappa_r \kappa_m)$ = 580 for a magnetic volume of 320 µm³, which is large, considering the small actual volume, compared to many other magnon-photon hybrid systems[3,31,32].

The linear increase of $g$ with the square root of the volume for a given center conductor width is in agreement with previous work, which showed that the coupling strength between the two systems can be described by[31,42]

$$g = g_e \mu_B b_{rf} \omega_r \sqrt{\frac{N}{8\hbar Z}} = g_s \sqrt{N} \qquad (3)$$

where $b_{rf} \equiv B_{rf}/I$ the magnetic induction per unit current, $Z$ is the characteristic impedance of the resonator, $g_e \approx 2$ is the electron gfactor, $\mu_B$ is the Bohr magneton, $\hbar$ is the reduced Planck constant, and $g_s$ is the single spin coupling rate. The magnitude of $b_{rf}$ for a CPW can be approximated[43] using the Karlqvist equation as $b_{rf} \approx \mu_0/(2W_{cc})\Gamma(y, W_{cc})$, where $\Gamma = (2/\pi)tan^{-1}(W_{cc}/(2y))$ is a spacing loss, and is very close to 1 for the entire thickness of the magnetic film. Using this relation and Eq. 3, we estimate $g_s/(2\pi)$ = 15 Hz for $W_{cc}$ = 50 µm and 74 Hz for $W_{cc}$ = 10 µm. The single spin coupling strength can be experimentally obtained from the slope of a linear fit to the data in *Figure 3* and results in $g_s/(2\pi)$ = 14.6 Hz ± 0.4 Hz for $W_{cc}$ = 50 µm and 66.8 Hz ± 6.2 Hz for $W_{cc}$ = 10 µm. The experimentally determined $g_s$ for $W_{cc}$ = 50 µm is within 2% of that predicted by Eq. 3, while the discrepancy is around 10% for $W_{cc}$ = 10 µm. The data show an increase in single spin coupling by a factor of 4.6 by decreasing $W_{cc}$ from 50 µm to 10 µm, compared to the factor of 5 predicted by Eq. 3.

To directly measure the change in coupling with center conductor width and to investigate the slightly reduced single-spin coupling for the 10 µm width resonator, a series of resonators with identical nominal frequencies and impedances (13 GHz and 50 Ω, respectively) but varying $W_{cc}$ were fabricated. Each resonator was loaded with a magnetic element of the same nominal volume (192 µm³) and patterned as mentioned above. The measured coupling strength vs. $1/W_{cc}$ are shown in Figure 4 and shows generally shows good agreement with the trend predicted by Eq. 3 using the Karlqvist equation with the exception of the resonator with $W_{cc}$ = 6 µm that has coupling reduced by 10% compared to theory, similar to the result for $W_{cc}$ = 10 µm in *Figure 3*. Even with the reduced coupling compared to theory, the $W_{cc}$ = 6 µm resonator has a coupling strength of 707 MHz ± 11 MHz and $\kappa_m/(2\pi)$ = 218 MHz ± 4.5MHz, resulting in a cooperativity, C = 720. This is corresponds to a 25% increase in cooperativity compared to that obtained for $W_{cc}$ = 10 µm in Fig. 3, despite employing a magnetic element with 40 % smaller volume, indicating that reducing center conductor width is an effective means of increasing cooperativity without the need to increase the volume of magnetic material.

The deviation of the data from that predicted by the Karlqvist approximation in Eq. 3 can likely be explained by looking at the magnetic field profile generated by a superconducting current distribution and the specific sample geometry used in this experiment. For this geometry, the current density in the center conductor can be approximated as[44] $J(x) = J(0)(1 - (2x/W_{cc})^2)^{-1/2}$, where $J(0)$ is the current density at the center of the waveguide, and $x$ is the lateral position with $x = 0$ at the center of the waveguide. The resultant in-plane magnetic fields for both a superconducting current distribution and a uniform distribution (normal metal at low frequency) are plotted in *Figure 5* (a) and (b) for $W_{cc}$ = 50 µm and 6 µm respectively at a height of 20 nm above the center conductor. Also plotted are the extents of the center conductor as well as the ferromagnet that (as mentioned above) was patterned 2 µm narrower than the center conductor to ensure reliable overlay. The current crowding at the edge of the superconducting line results in an increased magnetic field at the edge and a suppressed field in the center compared to the uniform distribution. The roll-off in the field strength from the superconductor edge becomes more rapid with decreasing conductor width, resulting in a smaller mean magnetic field seen by the ferromagnetic element on a superconducting line compared to a non-superconducting line. This, combined with ferromagnet being 2 µm narrower than $W_{cc}$, results in a suppression of the coupling compared to that expected from a resistive waveguide.

However, whereas the mean field compared to the uniform distribution is decreasing for narrower waveguides, the absolute magnitude of the of the field is also increasing, just as with the Karlqvist approximation. The sum of both of these effects is shown by plotting the mean field, which is directionally proportional to coupling strength, versus $1/W_{cc}$ in *Figure 5*(c). Here we can see that trend in the experimental data is well reproduced with narrower resonators showing decreased coupling compared to a linear fit. Also plotted is the expected coupling strength for the uniform current distribution. It appears that, in general, going to a narrower center conductor results in lower coupling for the superconducting case. However, if the center conductor width is made to be on the order of the London penetration depth, then the non-uniformity of the current density decreases, and the coupling is nearly identical for both superconducting and normal metal samples. Clearly, the use of the Karlqvist equation is useful for estimating how coupling scales with center conductor width, but a more detailed analysis is necessary to predict quantitative changes in coupling when the device is scaled, especially approaching length scales on the order of the London penetration depth. Such an analysis should consider the superconducting current density in the conductor and the specific geometry of the resonator and ferromagnet, as well as all components of the RF field and the ferromagnetic AC susceptibility.

The coupling in Eq. 3 is also predicted to increase linearly with resonator frequency, but it is not clear if this also applies to higher order harmonics, or only the fundamental. To investigate this, we looked at the coupling as a function of frequency in two ways. In one set of experiments, five resonators were fabricated with decreasing physical length to span a frequency range from 6 GHz to 30 GHz. Additionally, the harmonics of the longest 6 GHz resonator were probed. The dimension of the ferromagnetic element was 250 µm x 48 µm x 40 nm for all resonators.

This data are summarized in Figure 6 for both the harmonic response and the physically separate resonators. The data for the odd harmonics of the 6 GHz resonator are well fit by a linear approximation. The even harmonics demonstrate very low coupling due the presence of a $b_{rf}$ node at the center of the resonator where the magnetic element is located. The trend for the resonators with different physical lengths is monotonic and we can see that the first three frequencies are well represented by a linear approximation. The spectra from the two highest frequency resonators (24 GHz and 30 GHz) show coupling to many higher order spin wave modes in the ferromagnet resulting in a smearing of the upper branch of the spectrum. This made fitting of the data unreliable and they are not plotted in Figure 6. It is interesting to note the different slopes for using the harmonics versus physical length to vary the frequency. The predicted trend of coupling strength as a function of resonator frequency from Eq. 3 is 11.4 MHz/GHz which is in reasonable agreement with a linear fit to the fundamental harmonic resonators with a slope of 13.6 MHz/GHz. The odd harmonics show less efficient coupling at higher frequencies and the coupling strength increases with mode frequency by 2.8 MHz/GHz. The reduced coupling, while still maintaining a linear trend for the odd harmonics, is not immediately obvious given that $Z$ should be similar for all harmonics[45] and the magnetic element is still situated at an antinode of $b_{rf}$, with the magnitude determined by geometry, independent of frequency. Regardless of the underlying mechanism, these results demonstrate that for half-wavelength resonators, fabricating a physically shorter resonator results in higher coupling efficiency compared to operating at a higher odd harmonic.

The above results demonstrate that hybrid systems utilizing superconducting CPW resonators and the low-damping, high-moment, $Fe_{75}Co_{25}$ ferromagnet show potential for scaling to smaller magnetic volumes, higher coupling strengths, and larger cooperativities. We demonstrated a coupling strength of 707 MHz for a 13 GHz resonator ($g/\omega_r$ = 0.05), approaching the ultrastrong coupling regime where $g/\omega_r$ > 0.1. Our results suggest that for a given magnetic volume, by slightly decreasing the center conductor width to 3 µm we would be in ultrastrong coupling regime, and for $W_{cc}$ around 200 nm, we would be able to enter the deep strong coupling regime where the coupling strength exceeds the resonator frequency. This assumes that we extend the length (or increase the thickness) of the ferromagnet to maintain constant magnetic volume. Even without this assumption the fact that $g \propto \sqrt{V_{mag}}$ and $g \propto 1/W_{cc}$ leads to $g \propto W_{cc}^{-1/2}$. Therefore, in our system if we simply reduce $W_{cc}$ while maintaining constant length and thickness, we would expect to achieve ultrastrong coupling for $W_{cc}$ around 2 µm and deep strong coupling for resonators with center conductors less than 20 nm. Additionally, it may be possible to utilize other resonator geometries such as lumped-element inductive-capacitive (LC) resonators with lower impedance and extremely narrow inductive constrictions to further increase coupling.

This system has the potential ability to further reduce damping by improving the design of the hybrid system. In the present experiments the ferromagnetic stack was optimized for room temperature operation. It is reasonable to assume that small changes in the specific Fe-Co alloy composition or different seed/cap could result in larger coupling and/or lower damping at 4 K. Given the flexibility of microfabricated devices and the ease of performing resonant and broadband measurements in a single cooldown, our system can be used to rapidly evaluate different ferromagnetic stacks to optimize performance at 4 K and below. Additionally, devices with anti-damping torques can be engineered, with the goal of achieving performance on par with ferromagnetic insulators. The flexibility of microfabricated designs also enables straightforward coupling of multiple magnetic elements, potentially to multiple resonators, with the ability to apply fast rf-magnetic pulses to explore the time response of these systems in the strong coupling regime.

In conclusion, we investigated the coupling between the ultra-low damping metallic ferromagnetic alloy $Fe_{75}Co_{25}$ alloy and various superconducting CPW resonators and demonstrated coupling strengths over 700 MHz for a 13 GHz resonator. The role of the width of the center conductor was investigated and it was shown that reducing the width of the center conductor is an effective way to increase coupling even when the ferromagnetic volume is reduced. It was also demonstrated that changing the physical dimension of the resonator if more effective at increasing coupling compared to utilizing harmonics. All these results indicate the feasibility of the scaling of hybrid magnetic-microwave systems to smaller dimensions.

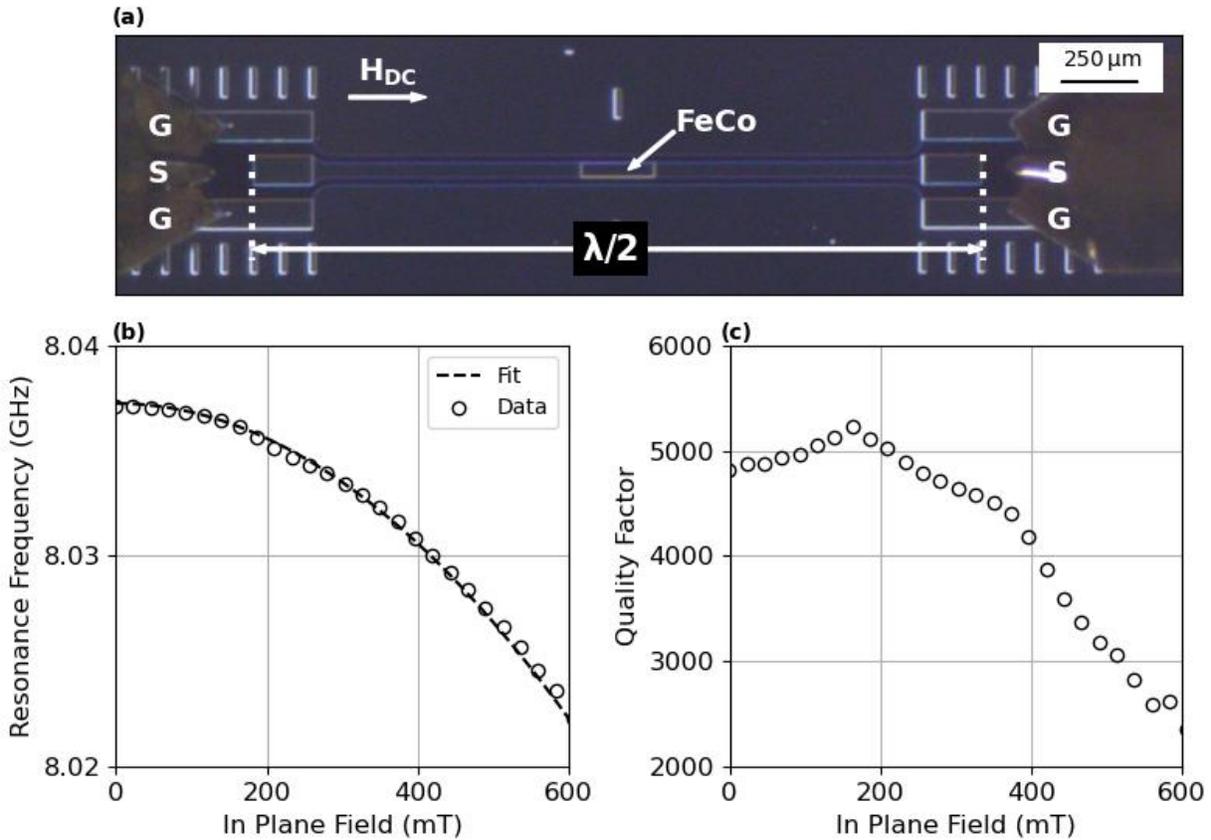

Figure 1: (a) Optical image of a typical device, including the GSG probes. (b) The response of a bare Nb resonator at 3.2 K to an applied in-plane field, showing the parabolic dependence of resonance frequency to applied field. (c) The resonator quality factor is above 4000 for fields below 400 mT, where the majority of the strong coupling was measured in this study.

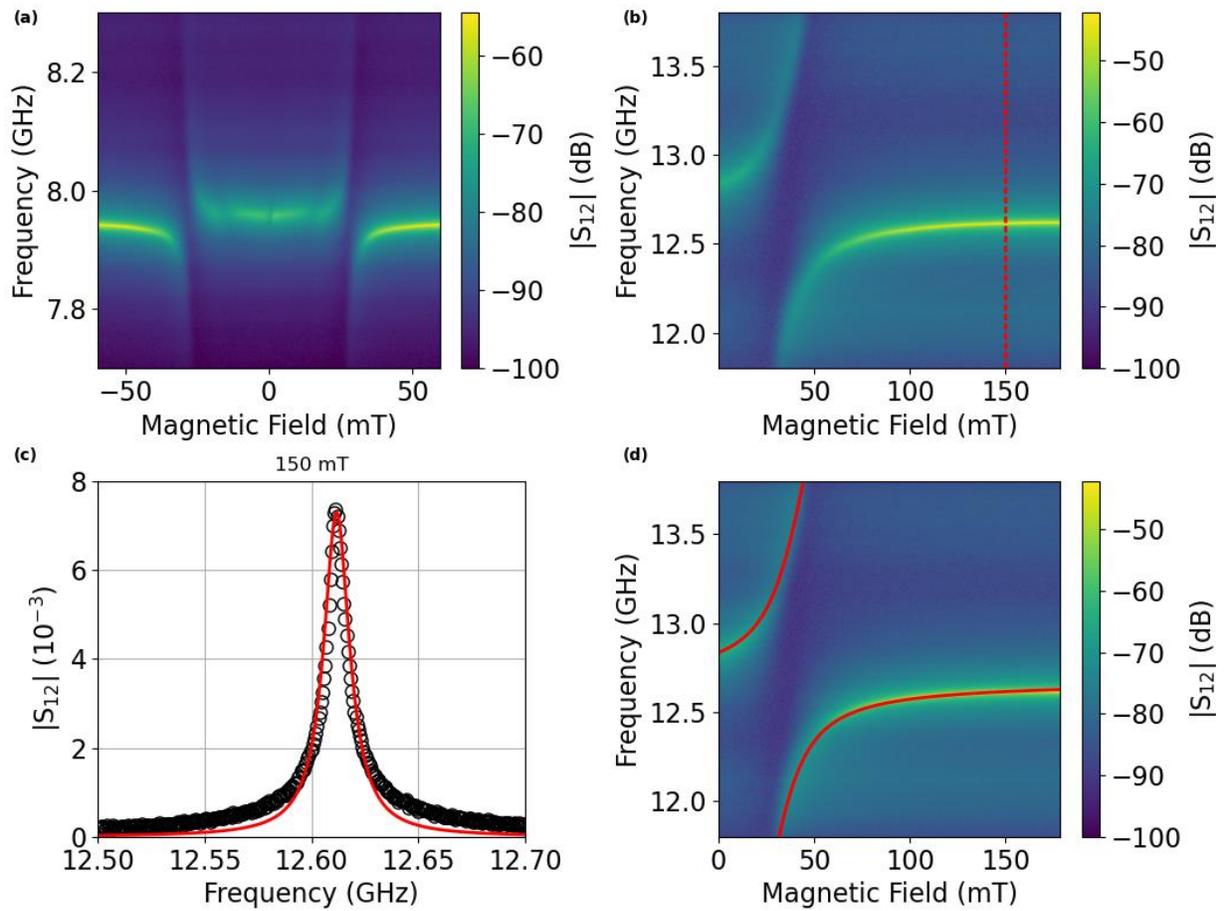

Figure 2: (a) The full spectral response of the hybrid system showing symmetry response for a magnetic volume of 480 µm$^3$ and $W_{cc}$ = 50 µm. The field was swept from negative to positive magnetic field. (b) Spectrum of a magnetic volume of 192 µm$^3$ with $W_{cc}$ = 6 µm. The dashed red line corresponds to the response at 150 mT plotted in (c). (d)The eigenvalue solution overlaid on the original spectrum showing good agreement between the fit and maximum transmission and a coupling strength of 707 MHz.

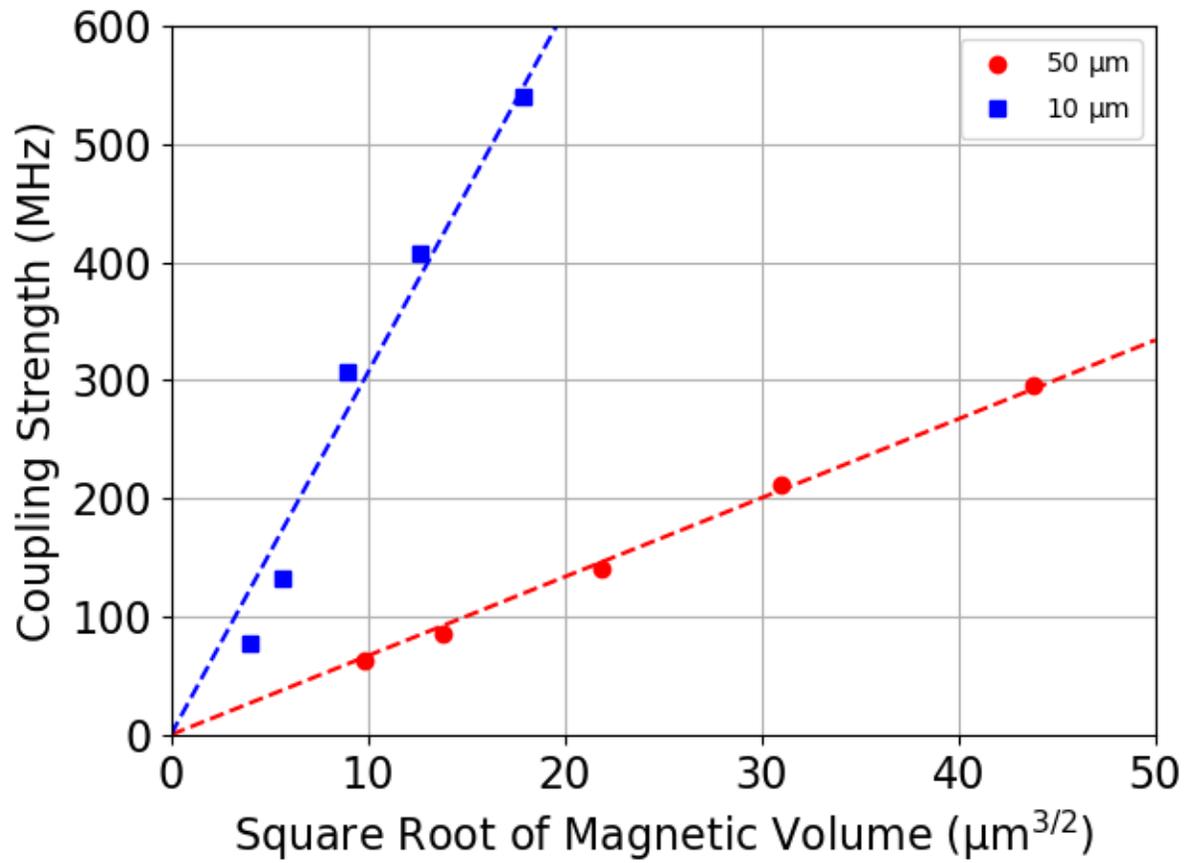

Figure 3: Coupling strength compared to the square root of the volume of the magnetic element. The slope is given by the single pin coupling strength, $g_s$.

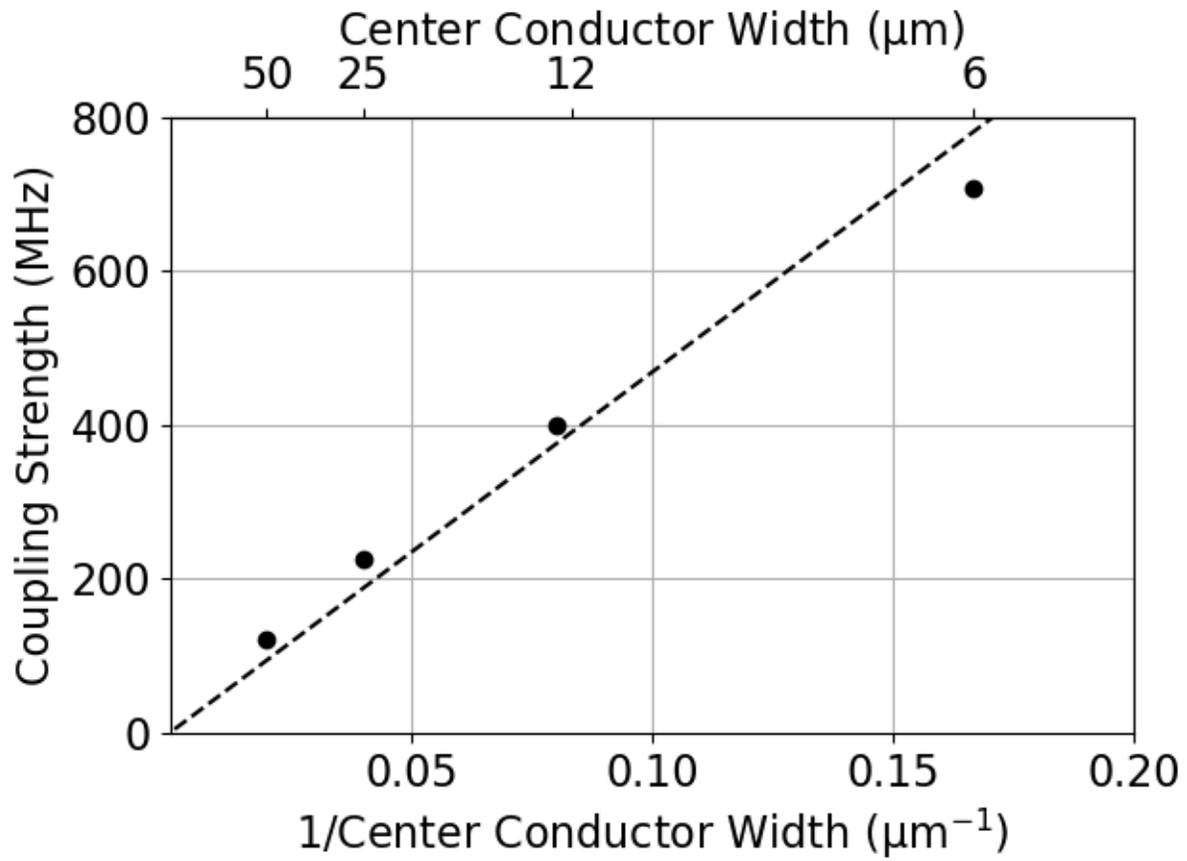

Figure 4 Effect of scaling the center conductor width on the coupling rate for a magnetic volume = 192 μm³. The dashed line is the predicted coupling strength given by Eq. 3.

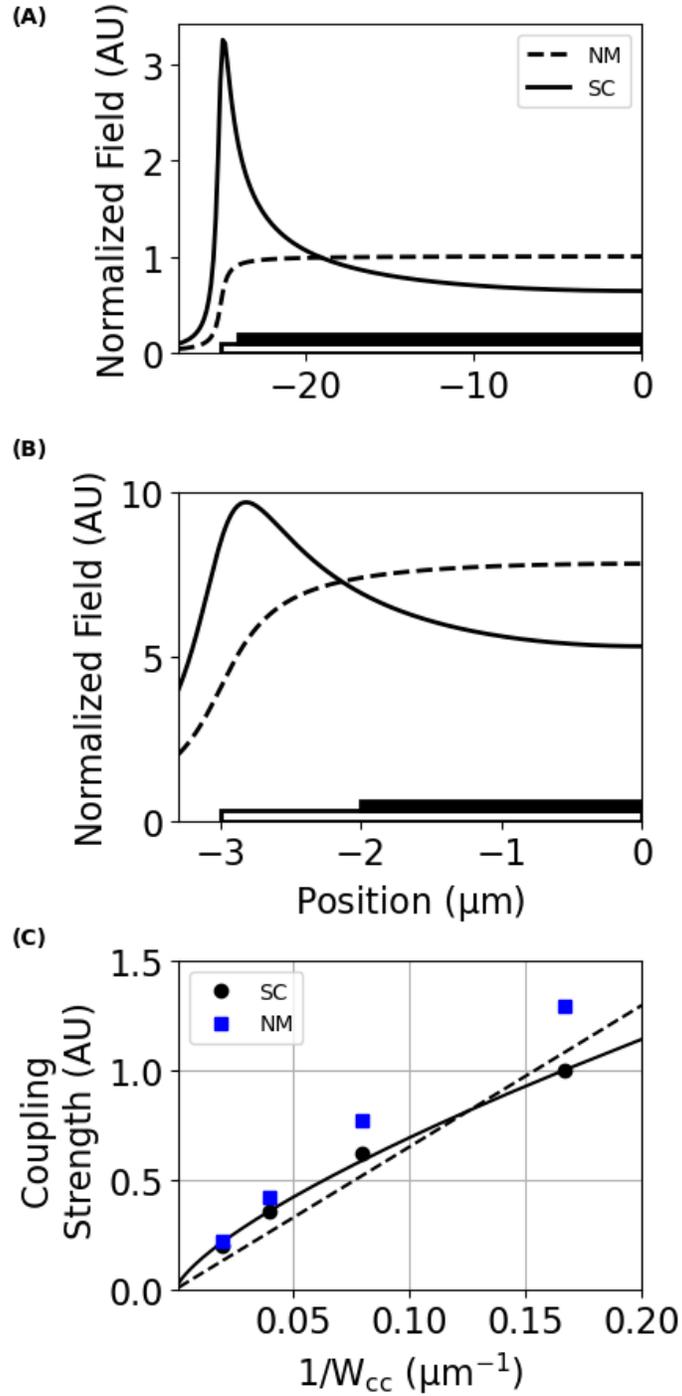

Figure 5: In-plane magnetic field profiles calculated from Ref. 44 for (a) $W_{cc}$ = 25 μm and (b) $W_{cc}$ = 6 μm with the field magnitude normalized to the center field of the 50 μm sample. The white rectangle at the bottom of the plot demonstrates the extent of the center conductor, and the black rectangle the ferromagnet for half of the mirror-symmetric CPW center conductor. For all samples, the magnetic material is 2 μm narrower than the center conductor resulting in narrower resonators seeing a lower mean magnetic field compared to a uniform current distribution. (C) the coupling strength for both superconducting (black circles) and uniform (blue squares) current distributions, calculated from Eq. 4.

The dashed line is a linear fit to the black circles. The simulated coupling of the resonators with the superconducting current distribution displays a power law dependence with k = -0.73 (solid black line), in excellent agreement with the experimental result.

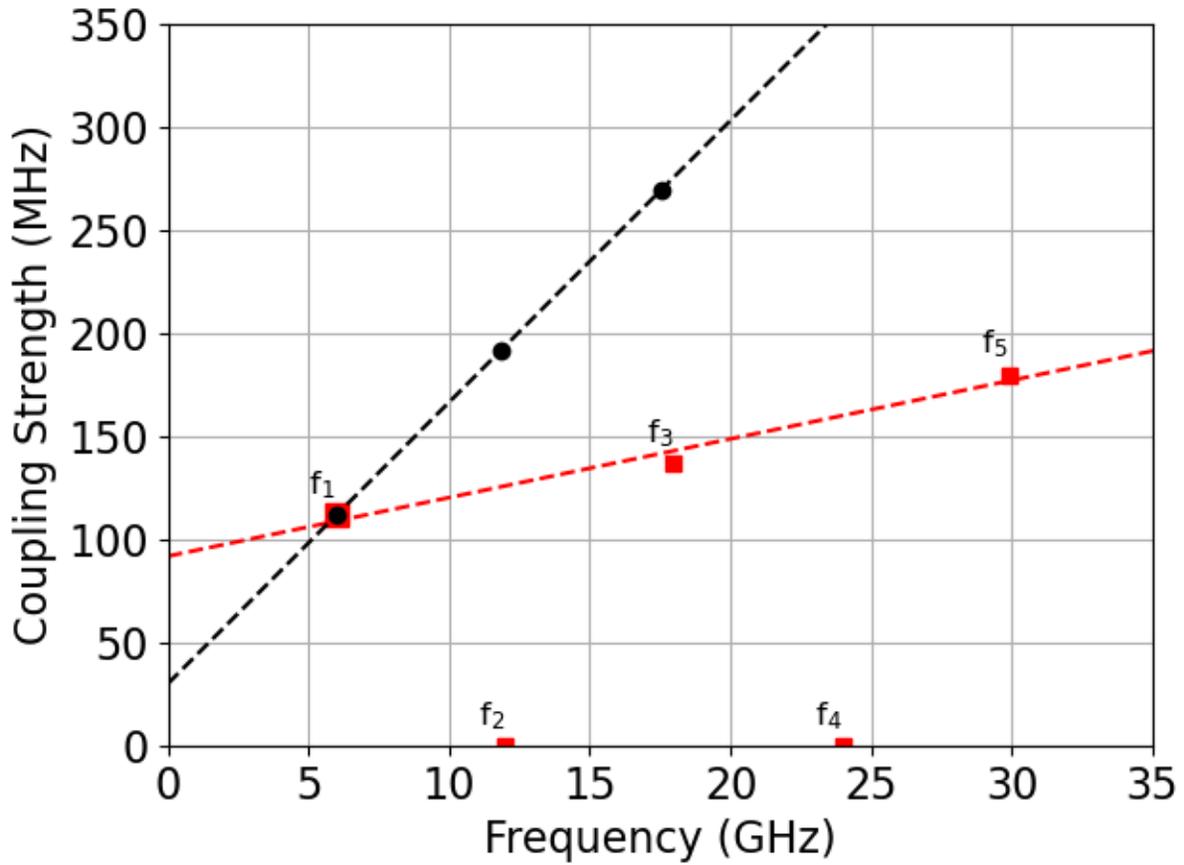

Figure 6: Frequency response of five resonators with different lengths (black circles) as well as the harmonics of the longest resonator (red squares). The black dashed line is a linear fit to the lowest three frequencies while the red dashed line is a fit for the fundamental, third, and fifth harmonics.